\documentclass[aps,amsfonts,preprint,nofootinbib]{revtex4}
\def\S2{\bar{S}}
\def\a{\alpha}
\def\b{\beta}

\def\and{a_{n}^\dagger}

\def\sn2d{\Sn2^\dagger}
\def\ad {{\dot\alpha}}

\def\bd {{\dot\beta}}

\def\({\left(}
\def\){\right)}
\def\<{\left\langle}
\def\>{\right\rangle}

\begin{document}
\title{Superspace Type II 4D Supergravity from Type II Superstring}

\author{Daniel. L. Nedel{\footnote{daniel@ift.unesp.br}}}

\affiliation{Instituto de F\'{\i}sica Te\'{o}rica, Unesp, Pamplona 145,
S\~{a}o Paulo, SP, 01405-900, Brazil }

\begin{abstract}

We derive the equations of motion of type II 4D
supergravity in superspace.  This is achieved by coupling the Type
II Berkovits' hybrid superstring to an $N=2$ curved background and
requiring that the sigma-model has N=(2,2) superconformal
invariance at one loop. We show that there are no anomalies in the
fermionic OPE's and the complete set of compensator's equations is
derived from the energy-momentum tensor. The equations of motion
describe a hypertensorial and vectorial multiplet coupled to a  $U(1)\times U(1)$  $N=2$  Poincar\`e Supergravity.
\end{abstract}

\maketitle


 


Superspace low-energy effective actions play an important role in the
study of string theory. They  provide important pieces of evidence for the existence of
various dual descriptions of string theories and permit an off-shell description  of the supergravity selected by string theory. Also, in the point of view of supergravity theories, the study of string corrections to the superspace equations of motion provides a new arena to develop futher superspace techniques, that can be very important to the development of supersymmetric theories \cite{Gates1}.

One way to construct low-energy effective actions in string
theory is looking for the low-energy equations of motion. This is
achieved by defining the sigma-model for the string in a curved background and  requesting conformal invariance \cite{Frad1}.

To derive the superspace equations of motion of the N=2
 supergravity theory that comes from  Type II
 superstring, we need to formulate the sigma-model directly in terms of a target superspace and which can be quantized covariantly. This is not the case of sigma models built with Green-Schwarz or RNS formalism. An alternative formalism for superstrings was  discovered by Berkovits
with local N=2 worldsheet superconformal invariance.

 This formalism is known
as hybrid formalism and is related to the RNS formalism by a
field-redefinition \cite{NB};
it is especially well-suited for compactifications to four
dimensions, where it allows manifestly $4D$ super-Poincar\'e
covariant quantization \cite{NB}.
The coupling of the theory to background fields was discussed in
\cite{NS}.
In \cite{boer}, the low-energy
effective equations of motion of heterotic superstring were
derived directly in superspace by means of this formalism. It was the first time that a kind of beta function calculation was made in a complete $4D$ supersymmetric way. 

In \cite{pure}
Berkovits started the development of  another formalism based on pure
spinors; it allows manifestly $10D$ super-Poincar\'e covariant
quantization.  In ref \cite{nahowe} the pure spinor formalism was used to construct a sigma model directly in  $10D$ curved superspace and by means of a tree level analysis of this sigma model, the structure of $10D$ superspace type II supergravity was derived. For the $4D$ case, the hybrid formalism was explored in \cite{eu} and the complete set of type II torsion's constraints was derived from the hybrid type II sigma model. In addition, a field redefinition
 enabled to get information of the dilaton at tree level. It was shown that this information is consistent with the compensator's superspace type II equations of motion.

 In this paper, a one loop analysis of the Type II hybrid sigma model is carried out and the
 superspace equations of motion for Type II supergravity are derived. The methods presented
here are a generalization of  the methods developed for the heterotic
superstring in \cite{boer}.
 
The core of the hybrid formalism lies in the fact that a
critical N=1 string can be formulated as a critical N=2 string,
without changing the physical content \cite{otha}. This is achieved by
twisting
the ghost sector of the critical N=1 string .
 After performing this embedding for the critical
RNS superstring, a field redefinition allows the resulting N=2
string to be made manifestly spacetime supersymmetric for
compactifications to four dimensions. In this case, the critical
$c=6$ matter sector splits into a $c=-3$ four-dimensional part and
a $c=9$ compactification-dependent part. The action for this
superstring can be written as : $S=S_{4D}+\a^{\prime}S_{FT}+S_c$,
where the  first term describes the classical four dimensional
part, the second one is the Fradkin-Tseytlin term which  has the
dilaton coupling and $S_c$ is the action for the
compactification-dependent $c=9$ theory. This work does not
 concern about the fields that depend on compactification, so it is
enough to concentrate just in the $N=(2,2)$ $c=-3$ sector.
 In a flat 4d background,
the type II superstring is described, in the $N=(2,2)$ superconformal gauge,
 by the following action:

\begin{eqnarray}
S &=&\frac{1}{\alpha ^{\prime }}\int d^{2}z\frac{1}{2}\partial x^{m}%
\overline{\partial }x_{m}+p_{\alpha }\overline{\partial }\theta ^{\alpha
}+p_{\dot{\alpha}}\overline{\partial }\theta ^{\dot{\alpha}}++\widehat{p}%
_{\alpha }\partial \widehat{\theta }^{\alpha }\nonumber \\
&+&\widehat{p}_{\dot{\alpha}
}\partial \widehat{\theta }^{\dot{\alpha}}-\frac{\alpha ^{\prime }}{2}%
\overline{\partial }\rho \left( \partial \rho +a_{z}\right)  
+\frac{\alpha ^{\prime }}{2}\partial \widehat{\rho }(\partial \widehat{%
\rho }+\widehat{a}_{z})+S_{c} + \a^{\prime}S_{FT}. \label{acao1}
\end{eqnarray}

The four-dimensional
 part of the action contains the spacetime
variables, $x^m$ ($m=0$ to 3), the left-moving fermionic
variables, $\theta^{\a}$ and $\theta^{\dot{\a}}$, 
the conjugate left-moving fermionic variables, $p_{\a}$ and $\bar
{p}_{\dot{\alpha}}$, and one left-moving boson, $\rho$, with a
`wrong' sign for the kinetic term. The right-sector of the Type II
superstring is described by the right-moving fermionic fields,
$\hat{\theta}{}^{\alpha}$,$\hat{\bar{\theta}}{}^{\dot{\alpha}}$,
the conjugate $\hat {p}_{\a}$, $\hat{ \bar p}_{\ad}$, and
one right-moving boson, $\hat{\rho}$. The fields $a_z$, $\hat
{a}_{\bar{z}}$ are the worldsheet $U(1)\times U(1)$ gauge fields
($e^\rho$ carries U(1) charge).
 The $c=-3$
$N=(2,2)$ superconformal currents are defined by the left and right
energy-momentum tensors, $T$ and  $\widehat{T}$;  the left and right
fermionic generators, $
G,\overline{G},\widehat{G}$ and $\widehat{\overline{G}}$; and the
$U(1)\times U(1)$ currents, $J$ and $\widehat{J}$.  The left moving ones can be written as


\begin{eqnarray}
T &=&\left( -\frac{1}{2}\Pi ^{\alpha \dot{\alpha}}\Pi _{\alpha \dot{\alpha}%
}-d_{\alpha }\partial \theta ^{\alpha }-d_{\dot{\alpha}}\partial \overline{%
\theta }^{\dot{\alpha}}+\frac{\alpha ^{\prime }}{2}\partial \rho \partial
\rho +\partial ^{2}\rho \right)\nonumber \\
G &=&\frac{1}{i\alpha ^{\prime }\sqrt{8\alpha ^{\prime }}}\exp \left(
i\rho \right) d^{\alpha }d_{\alpha }, \nonumber \\
 \overline{G}&=&\frac{1}{i\alpha ^{\prime }\sqrt{8\alpha ^{\prime }}}\exp
\left( -i\rho \right) d^{\dot{\alpha}}d_{\dot{\alpha}}  \nonumber \\
J &=&-i\partial \rho \nonumber \\
\label{gera}
\end{eqnarray}
where we have used Pauli matrices to write  vectors in terms of
bi-espinors and we have defined:
\begin{eqnarray}
d_{\alpha } &=&p_{\alpha }+i\overline{\theta }^{\dot{\alpha}}\partial
x_{\alpha \dot{\alpha}}+\frac{1}{2}\overline{\theta }^{2}\partial \theta
_{\alpha }-\frac{1}{4}\theta _{\alpha }\partial \left( \overline{\theta }%
\right) ^{2}  \nonumber \\
d_{\alpha } &=&p_{\dot{\alpha}}+i\overline{\theta }^{\alpha }\partial
x_{\alpha \dot{\alpha}}+\frac{1}{2}\theta ^{2}\partial \overline{\theta }%
_{\alpha }-\frac{1}{4}\overline{\theta }_{\alpha }\partial \left( \overline{%
\theta }\right) ^{2}.  \nonumber \\
\Pi _{a} &\rightarrow &\Pi _{\alpha \dot{\alpha}}=\partial x_{\alpha \dot{%
\alpha}}-\theta _{\alpha }\partial \overline{\theta }_{\dot{\alpha}%
}+i\partial \theta _{\alpha }\overline{\theta }_{\dot{\alpha}}.
\label{def}
\end{eqnarray}


The  right-moving $c=-3$ N=(2,2) generators are


\begin{eqnarray}
\widehat{G} &=&\frac{1}{i\alpha ^{\prime }\sqrt{8\alpha ^{\prime }}}e^{i%
\widehat{\rho }}\widehat{d}^{\alpha }\widehat{d}_{\alpha }
\  \nonumber \\
\widehat{\overline{G}} &=&\frac{1}{i\alpha ^{\prime }\sqrt{8\alpha ^{\prime }%
}}e^{-i\widehat{\rho }}\widehat{d}^{\dot{\alpha}}\widehat{d}_{\dot{\alpha}}
\nonumber \\
\widehat{T} &=&T=\left( -\frac{1}{2}\overline{\Pi }^{\alpha \dot{\alpha}}%
\overline{\Pi }_{\alpha \dot{\alpha}}-\widehat{d}_{\alpha }\overline{%
\partial }\widehat{\theta }^{\alpha }-\widehat{d}_{\dot{\alpha}}\overline{%
\partial }\overline{\widehat{\theta }}^{\dot{\alpha}}+\frac{\alpha ^{\prime }%
}{2}\overline{\partial }\widehat{\rho }\overline{\partial }\widehat{\rho }%
\right)  \label{anti}
\end{eqnarray}
where $\hat{ d}_{\alpha}$ and $\hat{\bar d}_{\dot{\alpha}}$ are
obtained from (\ref{def})  by using hatted variables and
replacing $\partial$ by $\bar{\partial}$. Using the free-field
OPE's the holomorphic (or left-moving) part of the $N=(2,2)$, $c=-3$
algebra can be written as
\begin{eqnarray}
T\left( z\right) T\left( w\right) &=&\frac{c/2}{\left( z-w\right) ^{4}}+%
\frac{2T\left( w\right) }{\left( z-w\right) ^{2}}+\frac{\partial T\left(
w\right) }{\left( z-w\right) },  \nonumber \\
T\left( z\right) G\left( w\right) &=&\frac{\frac{3}{2}G\left( w\right) }{%
\left( z-w\right) ^{2}}+\frac{\partial G\left( w\right) }{\left( z-w\right) }%
,  \qquad
T\left( z\right) \overline{G}\left( w\right) =\frac{\frac{3}{2}\overline{G}%
\left( w\right) }{\left( z-w\right) ^{2}}+\frac{\partial \overline{G}\left(
w\right) }{\left( z-w\right) },  \nonumber \\
J\left( z\right) G\left( w\right) &=&\frac{G\left( w\right) }{\left(
z-w\right) },  \qquad
J\left( z\right) \overline{G}\left( w\right) =\frac{\overline{G}\left(
w\right) }{\left( z-w\right) },  \nonumber \\
J\left( z\right) J\left( w\right) &=&\frac{c/3}{\left( z-w\right) ^{2}},
\qquad G\left( z\right) \overline{G}\left( w\right) =\frac{\frac{2}{3}c}{\left(
z-w\right) ^{3}}+\frac{2J\left( w\right) }{\left( z-w\right) ^{2}}+\frac{%
2T\left( w\right) +\partial J\left( w\right) }{\left( z-w\right) }.
\label{alge}
\end{eqnarray}


The anti-holomorphic (or right-moving) generators satisfy the same
algebra changing $(z,w)$ for $(\bar z,\bar w)$.
 In order to write a closed string action in target curved space one needs the vertex operators for the
  massless fields. For the type II superstring described in the hybrid formalism there is no
  distinction between Ramond and Neveu-Schwarz  vertex operators and both sectors are components of a superfield $U$ \cite{NS}. From the worldsheet point of view $U$ is an $N=(2,2)$ primary field of
conformal weight zero and $U(1)\times U(1)$ charge zero. 

The N=(2,2) primary field conditions and the gauge conditions
imply that $U$ is a prepotential for an N=2 conformal supergravity
coupled to a hyper-tensorial multiplet. The gauge fields of
supergravity sit in a Weyl multiplet with 24 bosonic and 24
fermionic off-shell components, while the matter fields are
described by a hypertensorial multiplet with  8 bosonic and 8
fermionic off-shell components. This prepotential represents the
massless compactification-independent fields of the Type II
superstring, without the dilaton. The dilaton does not couple
classically in the action; it is part of the compensator fields
and not part of the hypertensorial multiplet
. To know the precise form of the off-shell N=2
Poincar\`e supergravity that describes the low-energy effective
action for Type II superstrings, one needs to know the compensators
and the complete set of supergravity constraints, in particular
the torsion constraints that break the conformal invariance. In general, for $N=2$ supergravity it is necessary two compensators to pass from conformal to Poincar\'e \cite{gates},\cite{wit}.

From the $N=\(2,2\)$ worldsheet point of view the compensators  are selected by the $2D$ supersymmetry of the  Fradkin-Tseytlin term, that requires the following ones:  target chiral(anti-chiral) superfields $\phi_c (\overline{\phi}_c)$, that are related to a vector multiplet,
 and  target  twisted-chiral (anti-twisted-chiral) superfields  $\phi_{tc} (\overline{\phi}_{tc})$, that are related to a hypertensorial multiplet.

In an N=2 4d curved background , the type II superstring action is written in terms of the vielbein $E_M{}^A$. In the $N=(2,2)$ superconformal gauge the action is \cite{NS}:

\begin{eqnarray}
S &=&\frac{1}{\alpha ^{\prime }}\int d^{2}z\bigg[\frac{1}{2}\Pi ^{a}\overline{\Pi }_{a}+d_{\alpha }\overline{\Pi }%
^{\alpha _{+}}+d_{\dot{\alpha}}\overline{\Pi }^{\dot{\alpha}_{-}}+\widehat{d}%
_{\alpha }\Pi ^{\alpha _{-}}\nonumber \\
&+&\widehat{d}_{\dot{\alpha}}\Pi ^{\dot{\alpha}%
_{+}} 
+\frac{1}{2}\overline{\Pi }^{A}\Pi ^{B}B_{AB}+d_{\alpha }P^{\alpha \beta }%
\hat{d}_{\beta }+d_{\dot{\alpha}}P^{\dot{\alpha}\dot{\beta}}\hat{d}_{\dot{%
\beta}}
+ d_{\alpha }Q^{\alpha \dot{\beta}}\hat{d}_{\dot{\beta}}+d_{\dot{\alpha%
}}\bar{Q}^{\dot{\alpha}\beta }\hat{d}_{\beta }  \nonumber \\
&-&\frac{\alpha ^{\prime }}{2}\left( \overline{\partial }\rho +i\overline{%
\partial }\left( \phi _{c}-\overline{\phi }_{c}+\ \phi _{tc}-\overline{\phi }%
_{tc}\right) \right)
\times  \left( \partial \rho +i\partial \left( \phi _{c}-
\overline{\phi }_{c}+\phi _{tc}-\overline{\phi }_{tc}\right) +a_{z}\right)
\nonumber \\
&-&\frac{\alpha ^{\prime }}{2}\left( \partial \widehat{\rho }+i\partial
\left( \phi _{c}-\overline{\phi }_{c}-\ \phi _{tc}+\overline{\phi }
_{tc}\right) \right)
\times \left( \overline{\partial }\widehat{\rho }+i\overline{%
\partial }\left( \phi _{c}-\overline{\phi }_{c}-\ \phi _{tc}+\overline{\phi }%
_{tc}\right) +\widehat{a}_{z}\right) \bigg],  \label{sigmamodel}
\end{eqnarray}
where $\Pi^A=\partial Z^M E_M{}^A$ and $\overline{\Pi}^A =
\overline{\partial} Z^M E_M{}^A$. The coordinates of the N=2 target tangent superspace are  $Z^{A}\rightarrow \left( x^{a},\theta ^{\alpha_{+} },\theta ^{%
\dot{\alpha}_{-}},\theta ^{\alpha_{-} },\theta ^{\dot{\alpha}_{+}%
}\right)$ and  $ \pm$ is
an $SU(2)$-index.  
$P^{\alpha\dot{\beta}}$ and $Q^{\a\dot{\beta}}$ are chiral and
twisted-chiral field strengths of N=2 conformal supergravity whose
lowest components are the Type II Ramond-Ramond field strengths.
 Defining $\phi-\overline{\phi}= \phi
_{c}-\overline{\phi }_{c}+ \phi _{tc}-\overline{\phi }_{tc}$, the  holomorphic generators in curved superspace are:
\begin{eqnarray}
T&=&\frac{1}{\alpha ^{\prime }}\left[ -\frac{1}{2}\Pi ^{a 
}\Pi _{a}-d_{\alpha }\Pi ^{\alpha _{+}}-d_{%
\dot{\alpha}}\Pi ^{\dot{\alpha}_{-}}+\frac{\alpha ^{\prime }}{2}\left[\partial
\(\rho+\phi-\overline{\phi}\)\right]^2 \right]  
-\frac{1}{2}\partial ^{2}\left(\phi _{c}+\overline{\phi }_{c}+ \phi
_{tc}+\overline{\phi }_{tc} \right),\nonumber \\
G &=& \frac{1}{i\alpha ^{\prime }\sqrt{8\alpha
^{\prime }}}\exp \left(
i\rho \right) d^{\alpha }d_{\alpha}-\partial \left( \frac{1}{i\sqrt{2\alpha ^{\prime }}}
e^{i\rho }d^{\alpha }\nabla _{\alpha _{+}}\( \phi _{c}+\phi_{tc}\)\right), \nonumber\\
\overline{G}& =& \frac{1}{i\alpha ^{\prime }\sqrt{8\alpha
^{\prime }}}\exp \left(
-i\rho \right) d^{\dot\a }d_{\dot\a}-\partial \left( \frac{1}{i\sqrt{2\alpha ^{\prime }}}
e^{-i\rho }d^{\dot\a }\nabla _{\dot\a _{-}}\( \overline{\phi} _{c}
+ \overline{\phi}_{tc}\)\right)\nonumber \\
J &=& \partial\(-i{\rho}+ \phi - \overline{\phi }\) \label{ger}.
\end{eqnarray}
The anti-holomorphic ones are obtained from $\(\ref{ger}\)$  by using hatted variables and replacing $\partial$ by $\overline{\partial}$ and $\phi_c$ by $\phi_{tc}$. The $N=(2,2)$ algebra derived in (\ref{alge}) for Type II
superstring coupled to flat superspace must be satisfied in the
curved sigma-model. However, in curved space one no longer has
worldsheet fields satisfying free OPE's and a
perturbative approach to check the $N=(2,2)$ algebra is necessary. As usual in
string theory, $\alpha'$ counts the number of loops in the two-dimensional
quantum theory, but in the hybrid formalism the kinetic term of the bosons $\rho$ and $\hat{\rho}$
 does not have an explicit factor of $\frac{1}{\alpha'}$ in front of,
and therefore the $\alpha'$-perturbation theory does not make sense
for these fields.
This problem is solved by making the field redefinition:
$ \rho  \rightarrow \rho -i\left( \phi _{c}-\overline{\phi }_{c}+\ \phi
_{tc}-\overline{\phi }_{tc}\right) $ and
 $\ \widehat{\rho } \rightarrow \widehat{\rho }-i\left( \phi _{c}-\overline{%
\phi }_{c}-\phi _{tc}+\overline{\phi }_{tc}\right) $. After that, 
 $\rho$ and $\hat{\rho}$
obey the same free field OPE's that we have used to derive the
algebra N=2 in (\ref{alge}); for the other fields
perturbation theory will be used. Although this field redefinition permits a
well defined $\a^{\prime}$ perturbation theory, the
dependence of the fermionic currents on $\rho$ and
$\hat{\rho}$
results in a new tree level dependence of the generators on  $\phi_c$ and $\phi_{tc}$. Owing to this dependence  the sigma model $\(\ref{sigmamodel}\)$ is only consistent at tree level if the dilaton fields satisfy the following equations \cite{eu}:
\begin{eqnarray}
\nabla _{\widehat{\gamma }}\left( \phi _{c}-\overline{\phi }_{c}+\
\phi
_{tc}-\overline{\phi }_{tc}\right) &=&0,
\qquad
\nabla _{a}\left( \phi _{c}-\overline{\phi }_{c}+\ \phi
_{tc}-\overline{\phi
}_{tc}\right) =0,  \nonumber \\
\nabla _{\widetilde{\gamma} }\left( \phi _{c}-\overline{\phi }_{c}-\ \phi _{tc}+%
\overline{\phi }_{tc}\right) &=&0,
\qquad \nabla _{a}\left( \phi _{c}-\overline{\phi }_{c}-\ \phi
_{tc}+\overline{\phi }_{tc}\right) =0,  \label{eqarvore}
\end{eqnarray}
where was used: $T^{\widetilde{\a}}= \(T^{\dot{\a}_{-}},T^{\a_{+}}\)$ and $ T^{\widehat{\a}}=\(T^{\dot{\a}_{+}},T^{\a_{-}}\)$. Surprisingly, these equations are the Type II compensator's equations that are going to be derived here from one loop analysis. In $\(\ref{eqarvore}\)$ it was used  the covariant
derivative in the tangent superspace defined as:
\begin{equation}
\nabla _{A} =E_{A}{}^{M}\partial _{M}+\omega _{A\beta _{+}}{}^{\gamma
_{+}}M_{\gamma _{+}}{}^{\beta _{+}}+\omega _{A\dot{\beta}_{-}}{ }^{\dot{%
\gamma}}M_{\dot{\gamma}_{-}}{}^{\dot{\beta}_{-}} 
+\omega _{A\beta _{-}}{}^{\gamma _{-}}M_{\gamma _{-}}{}^{\beta
_{-}}+\omega _{A\dot{\beta}_{+}}{ }^{\dot{\gamma}_{+}}M_{\dot{\gamma}%
_{+}}{}^{\dot{\beta}_{+}}+\Gamma _{A}Y+\widehat{\Gamma }_{A}\widehat{Y}\nonumber, 
\label{dcovariant}
\end{equation}
where $\omega,\Gamma, \widehat{\Gamma}$ are the Lorentz and
$U(1)\times U(1)$ connections, $M$ are the Lorentz generators and
$Y,\widehat{Y}$ the $U(1)\times U(1)$ generators. It must be observed
that there are two independent spacetime spinors in the Type II
sigma-model, so one has two independent fermionic structure
groups. Thus, just like the two independent U(1) connections one
has two independent sets of irreducible spin connections:
$\omega_{A\a_+}{}^{\b_+}$, $\omega_{A\ad_{-}}{}^{\bd_-}$ and
$\omega_{A\a_{-}}{}^{\b_-}$, $\omega_{A\dot\alpha_{+}}{}^{\bd_+}$.
The covariant derivative satisfies the algebra
\begin{equation}
\left[ \nabla _{C,}\nabla _{A}\right \} =T_{CA}{}^{B}\nabla _{B}
+R_{CAE}{}^{D}M_{D}{}^{E}+F_{CA}Y+\widehat{F}_{CA}\widehat{Y},
\end{equation}
where 
 $F$ and $\widehat{F}$ are the super $U(1)\times U(1)$ field
strengths and $T$, $R$ are the supertorsions and supercurvatures. It was shown in \cite{eu} that the N=(2,2)
structure of the hybrid formalism selects a different version of the N=2 Poincar\'e Supergravities
described in \cite{wit}. For the type II superstring the ${SU(2)}/{U\(1\)}$ is fixed by the matter (hypermultiplet)
 and the
compensators are dynamical (string gauge). Part of the hypermultiplet that is not
fixed goes to the supergravity multiplet which, after imposing the
conventional constraints, presents
 $32+32$ off-shell degrees of freedom. At the end we have an $ N=2$ $U\(1\)\times U\(1\)$ Poincar\`e supergravity.


Next, we are going to analyse the sigma model at one loop. The
strategy is the same as in \cite{boer}. A typical beta-function
calculation does not guarantee the full $N=(2,2)$ superconformal
invariance. The latter would only follow from a standard
supersymmetric beta function calculation if the model could be
formulated in $N=(2,2)$  superspace on the worldsheet, which does
not seem possible. So, we need to check the $N=(2,2)$ algebra by
calculating the OPE's perturbatively. At tree-level, there are no
double contractions and it is necessary just to verify  the part
of the N=(2,2) algebra that depends on simple
contractions. At one loop one has double contractions. However, as the part of the generators that comes from the Fradkin-Tseytlin term has already an $\a^{\prime}$ dependence, one needs just to evaluate tree level OPE's when these generators are involved. To perform the perturbative check of the OPE's $(\ref{alge})$ in a covariant way, we are going to use a background covariant expansion that
preserves manifestly all the local symmetries of the target
superspace. The traditional way to achieve such an expansion is to use a
tangent vector that relates two points in target superspace, the
classical field and the quantum fluctuations, then expand all the
tensors in powers of this vector and use Riemann normal
coordinates to covariantize the expansion \cite{mukhi}. Each tensor $T$ of the action is expanded as follows:
 $ \Delta T=\left[ y^{A}\nabla _{A},T\right] $, 
where $y^{A}$ is the quantum flutuation. By applying iteratively the operator $\Delta$ we get a power series in  $y^{A}$ and classical fields. For example:

\begin{eqnarray}
\Delta \Pi ^{A}&=&\nabla y^{A}-\Pi ^{B}y^{C}T_{CB}{}^{A}, \nonumber \\
\Delta \left(\Pi^B \nabla_{B} y^{A}\right) &=&-Y^{D}\Pi ^{B}y^{C}R_{CBD}{}^{A}-\omega
\left( A\right) y^{A}\Pi ^{B}y^{C}F_{CB}-\widehat{\omega }\left( A\right)
y^{A}\Pi ^{B}y^{C}\widehat{F}_{CB}.
\end{eqnarray}
where $\nabla=\Pi^A\nabla_A$. Here $\Pi^A $ is a classical field  and $\omega\(A\)$, $\widehat{\omega}\(A\)$ are the $ U\(1\)\times U\(1\)$ weights of the index A. The only ones different from zero are:
$\omega(\alpha_{+})= \hat{\omega}(\alpha_{+})
=\frac{1}{2}$,$\omega(\dot\alpha_{-})=
\hat{\omega}(\dot\alpha_{+})= -\frac{1}{2}$. For the fermionic fields we choose the expansion: $ d_\a = d_\a + D_\a$ ( $ D_\a$  being a classical field) and the same for $\hat{d}_\a$. 
Now, we can describe the kind of calculation we intend to do. We need to expand the action up to order three in the quantum fields and two for the classical fields.  The result is:

\begin{eqnarray}
S^{2} &=&\frac{1}{2}\nabla y^{a}\overline{\nabla }y^{a}+d_{\alpha }\overline{%
\nabla }y^{\alpha +}+d_{\dot{\alpha}}\overline{\nabla }y^{\dot{\alpha}-}+%
\hat{d}_{\alpha }\nabla y^{\alpha -}+\hat{d}_{\dot{\alpha}}\nabla y^{\dot{%
\alpha}+}  \nonumber \\
&&\frac{1}{2}\overline{\nabla }y^{a}y^{C}\left( \Pi ^{B}T_{BC}{}^{a}\right) +%
\frac{1}{2}\nabla y^{a}y^{C}\left( \overline{\Pi}^{B}T_{BC}{}^{a}%
\right) -\frac{1}{4}\overline{\nabla }y^{C}y^{B}\left( \Pi
^{a}T_{BC}{}^{a}+2\Pi ^{A}H_{ABC}\right)   \nonumber \\
&&+\frac{1}{2}\overline{\nabla }y^{C}y^{B}\left( T_{BC}{}^{\tilde{\alpha}}D_{%
\tilde{\alpha}}\right) +\frac{1}{2}\nabla y^{C}y^{B}\left( T_{BC}{}^{\hat{%
\alpha}}\hat{D}_{\hat{\alpha}}\right) +d_{\tilde{\alpha}}y^{C}\left( 
\overline{\Pi}^{B}T_{BC}{}^{\tilde{\alpha}}\right) +\hat{d}_{\hat{%
\alpha}}y^{C}\left( \Pi ^{B}T_{BC}{}^{\hat{\alpha}}\right)   \nonumber \\
&&-\frac{1}{4}\nabla y^{C}y^{B}\left( \overline{\Pi }^{a}T_{BC}{}^{a}-2%
\overline{\Pi}^{A}H_{ABC}\right)   \nonumber \\
&&+\frac{1}{4}y^{B}y^{C}\left[ \left( \overline{\Pi }^{D}\overline{\Pi}%
^{a}-\overline{\Pi}^{D}\overline{\Pi }^{a}\right) T_{DCB}{}^{a}-2%
\overline{\Pi}^{D}T_{DCB}{}^{\tilde{\alpha}}D_{\tilde{\alpha}}-2\Pi
^{D}T_{DCB}{}^{\tilde{\alpha}}D_{\tilde{\alpha}}\right.   \nonumber \\
&&\left. +\overline{\Pi}^{D}\left( \left( -1\right) ^{E\left(
D+B\right) +CD}T_{C}{}^{Ea}T_{DB}{}^{a}+H_{DCB}{}^{E}\right) \Pi _{E}\right] 
\nonumber \\
&&+d_{\alpha }P^{\alpha \beta }\hat{d}_{\beta }+d_{\dot{\alpha}}P^{\dot{%
\alpha}\dot{\beta}}\hat{d}_{\dot{\beta}}+d_{\alpha }Q^{\alpha \dot{\beta}}%
\hat{d}_{\dot{\beta}}+d_{\dot{\alpha}}\bar{Q}^{\dot{\alpha}\beta }\hat{d}%
_{\beta }  \nonumber \\
&&+d_{\alpha }y^{A}\nabla _{A}P^{\alpha \beta }\hat{D}_{\beta }+d_{\dot{%
\alpha}}y^{A}\nabla _{A}P^{\dot{\alpha}\dot{\beta}}\hat{D}_{\dot{\beta}%
}+d_{\alpha }y^{A}\nabla _{A}Q^{\alpha \dot{\beta}}\hat{D}_{\dot{\beta}}+d_{%
\dot{\alpha}}y^{A}\nabla _{A}\bar{Q}^{\dot{\alpha}\beta }\hat{D}_{\beta } 
\nonumber \\
&&D_{\alpha }y^{A}\nabla _{A}P^{\alpha \beta }\hat{d}_{\beta }+D_{\dot{\alpha%
}}y^{A}\nabla _{A}P^{\dot{\alpha}\dot{\beta}}\hat{d}_{\dot{\beta}}+D_{\alpha
}y^{A}\nabla _{A}Q^{\alpha \dot{\beta}}\hat{d}_{\dot{\beta}}+D_{\dot{\alpha}%
}y^{A}\nabla _{A}\bar{Q}^{\dot{\alpha}\beta }\hat{d}_{\beta }  \nonumber \\
&&+D_{\alpha }y^{A}y^{B}\nabla _{B}\nabla _{A}P^{\alpha \beta }\hat{D}%
_{\beta }+D_{\dot{\alpha}}y^{A}y^{B}\nabla _{B}\nabla _{A}P^{\dot{\alpha}%
\dot{\beta}}\hat{D}_{\dot{\beta}}+D_{\alpha }y^{A}y^{B}\nabla _{B}\nabla
_{A}Q^{\alpha \dot{\beta}}\hat{D}_{\dot{\beta}}  \nonumber \\
&&+D_{\dot{\alpha}}y^{A}y^{B}\nabla _{B}\nabla _{A}\bar{Q}^{\dot{\alpha}%
\beta }\hat{D}_{\beta }  \label{s2}
\end{eqnarray}


\begin{eqnarray}
S^{3} &=&-\frac{1}{4}\left( \overline{\nabla }y^{a}\nabla y^{B}y^{C}+\nabla
y^{a}\overline{\nabla }y^{B}y^{C}\right) T_{CB}{}^{a}  \nonumber \\
&&+\frac{1}{3}\overline{\nabla }y^{A}\nabla y^{B}y^{C}H_{CBA}+\frac{1}{2}d_{%
\tilde{\alpha}}y^{B}\overline{\nabla }y^{C}T_{BC}{}^{\tilde{\alpha}}+\frac{1%
}{2}\hat{d}_{\hat{\alpha}}y^{B}\nabla y^{C}T_{BC}{}^{\hat{\alpha}}  \nonumber
\\
&&+\frac{1}{4}\left( \overline{\nabla }y^{a}\nabla y^{B}y^{C}\Pi ^{D}+%
\overline{\nabla }y^{a}y^{B}y^{C}\overline{\Pi }^{D}\right) T_{DCB}{}^{a}+%
\frac{1}{2}d_{\tilde{\alpha}}y^{B}y^{C}\Pi ^{D}T_{DCB}{}^{\tilde{\alpha}} 
\nonumber \\
&&+y^{B}y^{C}\overline{\nabla }y^{D}\left[ \frac{1}{12}\overline{\Pi }%
^{a}\left( T_{DCB}{}^{a}+\left( -1\right) ^{CD}\nabla
_{C}T_{DB}{}^{a}\right) \right.   \nonumber \\
&&\left. +\left( \frac{1}{6}T_{DC}{}^{E}H_{EB}{}^{A}-\frac{1}{4}%
T_{DC}{}^{a}T_{B}{}^{Aa}-\frac{1}{3}H_{DCB}{}^{A}\right) \overline{\Pi }%
_{A}\right]   \nonumber \\
&&+y^{B}y^{C}\overline{\nabla }y^{D}\left[ \frac{1}{12}\overline{\Pi }%
^{a}\left( T_{DCB}{}^{a}+\left( -1\right) ^{CD}\nabla
_{C}T_{DB}{}^{a}\right) \right.   \nonumber \\
&&\left. +\left( -\frac{1}{6}T_{DC}{}^{E}H_{EB}{}^{A}-\frac{1}{4}%
T_{DC}{}^{a}T_{B}{}^{Aa}+\frac{1}{3}H_{DCB}{}^{A}\right) \overline{\Pi }%
_{A}\right]   \nonumber \\
&&-\frac{1}{6}y^{B}y^{C}\overline{\nabla }y^{D}\left[ T_{DCB}{}^{\tilde{%
\alpha}}+\left( -1\right) ^{CD}\nabla _{C}T_{DB}{}^{^{\tilde{\alpha}%
}}\right] D_{^{\tilde{\alpha}}}  \nonumber \\
&&-\frac{1}{6}y^{B}y^{C}\nabla y^{D}\left[ T_{DCB}{}^{\hat{\alpha}}+\left(
-1\right) ^{CD}\nabla _{C}T_{DB}{}^{^{\tilde{\alpha}}}\right] \hat{D}_{^{%
\hat{\alpha}}}  \nonumber \\
&&+d_{\alpha }y^{A}\nabla _{A}P^{\alpha \beta }\hat{d}_{\beta }+d_{\dot{%
\alpha}}y^{A}\nabla _{A}P^{\dot{\alpha}\dot{\beta}}\hat{d}_{\dot{\beta}%
}+d_{\alpha }y^{A}\nabla _{A}Q^{\alpha \dot{\beta}}\hat{d}_{\dot{\beta}}+d_{%
\dot{\alpha}}y^{A}\nabla _{A}\bar{Q}^{\dot{\alpha}\beta }\hat{d}_{\beta } 
\nonumber \\
&&+d_{\dot{\alpha}}y^{A}\nabla _{A}\bar{Q}^{\dot{\alpha}\beta }\hat{d}%
_{\beta }  \nonumber \\
&&+\frac{1}{2}\left[ d_{\alpha }y^{A}y^{B}\nabla _{B}\nabla _{A}P^{\alpha
\beta }\hat{D}_{\beta }\right. +d_{\dot{\alpha}}y^{A}y^{B}\nabla _{B}\nabla
_{A}P^{\dot{\alpha}\dot{\beta}}\hat{D}_{\dot{\beta}}+d_{\alpha
}y^{A}y^{B}\nabla _{B}\nabla _{A}Q^{\alpha \dot{\beta}}\hat{D}_{\dot{\beta}}
\nonumber \\
&&+\left. d_{\dot{\alpha}}y^{A}y^{B}\nabla _{B}\nabla _{A}\bar{Q}^{\dot{%
\alpha}\beta }\hat{D}_{\beta }\right]   \nonumber \\
&&+\frac{1}{2}\left[ D_{\alpha }y^{A}y^{B}\nabla _{B}\nabla _{A}P^{\alpha
\beta }\hat{d}_{\beta }\right. +D_{\dot{\alpha}}y^{A}y^{B}\nabla _{B}\nabla
_{A}P^{\dot{\alpha}\dot{\beta}}\hat{d}_{\dot{\beta}}+D_{\alpha
}y^{A}y^{B}\nabla _{B}\nabla _{A}Q^{\alpha \dot{\beta}}\hat{d}_{\dot{\beta}}
\nonumber \\
&&+\left. D_{\dot{\alpha}}y^{A}y^{B}\nabla _{B}\nabla _{A}\bar{Q}^{\dot{%
\alpha}\beta }\hat{d}_{\beta }\right]   \nonumber \\
&&+\frac{1}{6}\left[ D_{\alpha }y^{C}y^{A}y^{B}\nabla _{B}\nabla _{A}\nabla
_{C}P^{\alpha \beta }\hat{D}_{\beta }\right. +D_{\dot{\alpha}%
}y^{C}y^{A}y^{B}\nabla _{B}\nabla _{A}\nabla _{C}P^{\dot{\alpha}\dot{\beta}}%
\hat{D}_{\dot{\beta}}  \nonumber \\
&&+D_{\alpha }y^{C}y^{A}y^{B}\nabla _{B}\nabla _{A}\nabla _{C}Q^{\alpha \dot{%
\beta}}\hat{D}_{\dot{\beta}}  \nonumber \\
&&+\left. D_{\dot{\alpha}}y^{C}y^{A}y^{B}\nabla _{B}\nabla _{A}\nabla _{C}%
\bar{Q}^{\dot{\alpha}\beta }\hat{D}_{\beta }\right]   \label{S3}
\end{eqnarray}
where:
\begin{eqnarray}
T_{DCB}{}^{A}&=&R_{DCB}{}^{A}+\omega \left( A\right) F_{DC}\delta _{B}{}^{A}+%
\widehat{\omega }\left( A\right) \widehat{F}_{DC}\delta
_{B}{}^{A}\nonumber \\
&+&T_{DC}{}^{E}T_{EB}{}^{A}+\left( -1\right) ^{CD}\nabla
_{C}T_{DB}{}^{A}  \label{tgrande}\nonumber\\
H_{DCB}{}^{A}&=&\nabla _{C}H_{DB}{}_{A}\left( -1\right)
^{CD}\nonumber \\
&-&T_{CA}{}^{E}H_{EDB}{}\left( -1\right) ^{A\left( B+D\right)
+CD}
+T_{DC}{}^{E}H_{EBA} \nonumber 
\end{eqnarray}

The kinetic part of the expanded action provides the worldsheet propagators. In
momentum space they are:

\begin{eqnarray}
\left\langle y^{a}\left( p\right) y^{b}\left( l\right) \right\rangle &=&\alpha
^{\prime }\eta^{ab}\delta \left( p+l\right) \frac{1}{\left| p\right| ^{2}}\nonumber 
\qquad
\left\langle d_{\alpha }\left( p\right) y^{\beta +}\left( l\right)
\right\rangle =\alpha ^{\prime }\delta_\alpha^\beta \delta \left( p+l\right) \frac{-ip}{\left|
p\right| ^{2}}\nonumber \\
\left\langle \widehat{d}_{\alpha }\left( p\right) y^{\beta -}\left( l\right)
\right\rangle &=&\alpha ^{\prime }\delta_\alpha^\beta \delta \left( p+l\right) \frac{-i\overline{p}
}{\left| p\right| ^{2}}. 
 \label{prop2}
\end{eqnarray}
The other part of
the action provides the vertices. By expanding the generators
using the same background covariant  expansion, the
 OPE's are calculated contracting the fields with the vertices in
the action. By demanding that the $N=(2,2)$ algebra is satisfied at one loop,  the equations of motion are derived for the background fields.

In order to avoid the problems with coordinate space
regularizations,  the OPE's are calculated in momentum space using
the following dimensional regularization:


\begin{eqnarray}
&&\int d^{d}p\frac{%
p^{a}\overline{p}^{b}}{[\left| P\right| ^{2}]^{\alpha }[\left| P-K\right|
]^{\beta }} = \nonumber \\
&&k^{a+1-\alpha -\beta }\overline{k}^{b+1-\alpha -\beta }\left| \frac{K^{2}}{%
\mu ^{2}}\right| ^{-\varepsilon }
 \times \sum_{i=0}^{i=a}\left(
\begin{array}{c}
a \\
i
\end{array}
\right) [\frac{\Gamma \left( 2-\alpha -\beta +b+i-\varepsilon \right) }{%
\Gamma \left( 2-2\varepsilon -\alpha -\beta +i+b\right) }\nonumber \\
&&\times \frac{\Gamma \left( \alpha +\beta -1-i+\varepsilon \right)}{\Gamma \left( 1+\varepsilon \right) \mu ^{-2\varepsilon }} \Gamma \left(
1-\varepsilon -\beta +i\right) ].
\end{eqnarray}
After that we go back to coordinate space by means of an inverse Fourier
transform. As we are calculating expectation values of conserved
currents, all the divergences are cancelled. The checking of all
OPE's of the $N=(2,2)$ algebra is a tedious and hard work.  
 Let's put just some results. For the $
G\(z\)G\(w\)$ OPE there is no  information at one loop. All the
terms are proportional to the constraints derived in \cite{eu}.
Also, we can show that there is no information coming from the
$G\(z\)\widehat{G}\(w\)$ OPE. For $G\(z\)\overline {G}\(w\)$ the
result is:
\begin{eqnarray}
\left\langle G\left( z\right) \overline{G}\left( w\right) \right\rangle &=&%
\frac{1}{\left( z-w\right) }[2T\left( w\right) -\frac{1}{8\a ^{\prime }}\left[D^{\alpha }R_{\dot{\beta}\alpha }{}^{\dot{\beta}\dot{\gamma}
}D_{\dot{\gamma}} + D^{\dot{\alpha}}R_{\beta\dot{\alpha}%
}{}^{\beta \gamma } 
D_{\gamma}  -\frac{1}{2}D^{\alpha }D^{\dot{\gamma}}F_{\dot{
\gamma}\a } 
\right] ].\nonumber \\
  \label{GGbar}
\end{eqnarray}
In this equation all the spinors have index $\a_i=\a_{+}$ and $\dot{\a}_{j}=
\dot\a_{-}$.
The first term is the right term for this part of the $N=\(2,2\)$
algebra. The second one  can be removed by redefining the expansion
for  $D^{\alpha }$. In general the anomalies that result in the
equations of motion come from anti-holomorphic terms in the
holomorphic OPE's and holomorphic terms in the anti-holomorphic
OPE's. The result above shows that there is no information from
fermionic currents. So, the equations of motion come from the
OPE's that involve the energy-momentum tensor. The result for $T\(z\)G\(w\)$ is:
 \begin{eqnarray}
\left\langle T\left( z\right) G\left( w\right) \right\rangle &=&\frac{\left(
\overline{z}-\overline{w}\right) }{\left( z-w\right) ^{3}}[\overline{\Pi }%
^{d}\left( R_{d\beta_{+}\gamma_{+} }{}^{\beta_{+} }+\frac{1}{2}F_{d\gamma_{+} }+\frac{1%
}{2}R_{\gamma_{+} b}{}_{d}{}^{b}-iT_{d\gamma_{+}\dot{\b_{-}}}{}^{\dot{\beta_{-}}%
} + 2\nabla _{d}\nabla _{\gamma_{+} }\left(  \phi
-\overline{\phi }\right)\right) D^{\gamma}\nonumber \\
 &+&\overline{\Pi }^{\widehat{\delta}}\left(R_{\widehat{\delta}\beta_{+}\gamma_{+}}{}^{\beta_{+}}+\frac{1}{2}F_{\widehat{\delta}\gamma_{+}}+2\nabla _{\widehat{\delta}}\nabla _{\gamma_{+} }\left( \phi -
\overline{\phi }\right)+ iT_{\widehat{\delta}\widehat{\b},\gamma_{+}}{}^{\widehat{\b}}  \right) D^{\gamma}+...] +\frac{3G}{2\( z-w\)^{2}}. \nonumber \\
 \label{tg}
\end{eqnarray}
Here the dots mean terms that are higher derivatives of the first one.
Now one has anti-holomorphic terms.  These are the anomalies and 
provide the equations of motion. The holomorphic term is the
right term for the $ N=\(2,2\)$ OPE's and we got undesirable holomorphic terms that were cancelled redefining the energy-momentum tensor by means of counter-terms. Also it was necessary to redefine the $D$ expansion. The
right expansion for the $D$ field is related to a shift symmetry in
the background expansion and may be interesting to show that the
selected expansion comes from Slavnov-Taylor identities. The results from $T\(z\)T\(w\)$ are not presented because it
will take many pages and all the information  about the
equations of motion can be read from $\(\ref{tg}\)$. The derived
equations are:

\begin{eqnarray}
R_{\widehat{\delta}\beta_{+} \a _{+}}{}^{\beta_{+} }+\frac{1}{2}F_{\widehat{\delta}
\alpha _{+}}+iT_{\widehat{\delta}\widehat{\b},\a_{+}}{}^{\widehat{\b}} &=&-2\nabla _{\widehat{\delta}}\nabla _{\alpha_{+} }\left( \phi -
\overline{\phi }\right)  \nonumber \\
R_{d\beta_{+}\gamma_{+} }{}^{\beta_{+} }+\frac{1}{2}F_{d\gamma_{+} }+\frac{1}{2}%
R_{\gamma_{+} b}{}_{d}{}^{b}-iT_{d\gamma_{+}\dot{\b_{-}}}{}^{\dot{\beta}_{-}}&=&2\nabla _{d}\nabla _{\gamma_{+} }\left( \phi -\overline{\phi }\right).
\label{finalTG}
\end{eqnarray}

From
Bianchi identities we have (with constraints derived in \cite{eu}) :
\begin{equation}
R_{\widehat{\delta}\beta_{+} \a _{+}}{}^{\beta_{+} }+ iT_{\widehat{\delta}\widehat{\b},\a_{+}}{}^{\widehat{\b}} =
 -\frac{5}{2} F_{\widehat{\delta}\alpha _{+}}.
\end{equation}
Using the fact that the combination: $ \exp\( \phi
_{c}-\overline{\phi }_{c}+\ \phi _{tc}-\overline{\phi }_{tc}\)$
has unitary $ U\(1\)$ charge with respect to generator $Y$, the
first equation can be written as:

\begin{equation}
\nabla _{\alpha _{+}}\nabla _{\widehat{\delta}}\left( \phi _{c}-\overline{\phi }%
_{c}+\ \phi _{tc}-\overline{\phi }_{tc}\right) =0.
\end{equation}
Making the same manipulations in the other equation one gets:
\begin{equation}
\nabla _{\alpha }\nabla _{b}\left( \phi _{c}-\overline{\phi }_{c}+\ \phi
_{tc}-\overline{\phi }_{tc}\right) =0.
\end{equation}

 The equations above are a derivative of the first line in $\(\ref{eqarvore}\)$.The same derivative of the second line is obtained from $\widehat{T}\widehat{G}$  OPE. 
  Although the derivatives of  an equation are a stronger condition than the  equation itself, there is no physical meaning to take these stronger conditions into account. So the right equations of motion for the type II compensators derived from the sigma model at one loop are the equations $\(\ref{eqarvore}\)$. These equations of motion describe the 16+16
degrees of freedom of a vector and a hypertensorial multiplet coupled to N=2
supergravity in the string gauge. It will be interesting to take
into account the compactification-dependent vertex operators. This
extension is straightforward and it will be possible to see how the
compactification will modify the 
super-geometry and the dynamics. The fact that there are no anomalies in the
fermionic OPE's means that a traditional beta function calculation
could provide the same equations of motion. There is no physical
explanation for this fact, but it will be interesting to check this
point in a two loop analysis.

\section*{Acknowledgments:} I would like 
to thank N. Berkovits, J. A. Helayel Neto, D. Z. Marchioro and M.C.B. Abdalla for useful discussions. The author is supported by a FAPESP
post-doc fellowship. This research is part of the author PhD thesis, advised by Nathan Berkovits and supported by FAPESP.


\begin{thebibliography}{99}

\bibitem{Gates1}
S.~J.~J.~Gates, A.~Kiss and W.~Merrell, {\it Dynamical equations from a first-order perturbative superspace formulation},
hep-th/0409104.
\bibitem{Frad1}E. Fradkin and A.
Tseytlin, {\it Quantum String Theory Effective Action, Phys. Lett.} {\bf B 158} (1985) 316.


\bibitem{NB}N. Berkovits, {\it The Ten-dimensional Green-Schwarz superstring is a twisted
Neveu-Schwarz-Ramond string, Nucl. Phys}. {\bf B 420}
(1994) 332 [hep-th/9308129]; N. Berkovits and C. Vafa, {\it On the Uniqueness of string theory,
Mod.\ Phys.\ Lett.} {\bf A 9} (1994) 653 [hep-th/9310170].

\bibitem{NS} N. Berkovits and W. Siegel, {\it Superspace Effective Actions for 4D Compactifications of Heterotic and Type
II Superstrings, Nucl. Phys.} {\bf B 462}
(1996) 213 [hep-th/9510106]. 


\bibitem{pure} N.~Berkovits,
{\it Super-Poincare covariant quantization of the superstring,
JHEP} {\bf 0004} (2000) 018 [hep-th/0001035].
\bibitem{nahowe} N.~Berkovits and P.~S.~Howe, {\it Ten-dimensional supergravity constraints from the pure spinor formalism  for the superstring, Nucl. Phys.} {\bf B 635} (2002) 75.

\bibitem{boer} J. de Boer and K. Skenderis, {\it Covariant computation of the low energy effective action of the  heterotic superstring, Nucl. Phys.} {\bf B 481} (1996) 129 [hep-th/9608078].
\bibitem{eu} Daniel.~L.~Nedel,
{\it Consistency of superspace low energy equations of motion of 4D Type II superstring with Type II sigma model at tree-level, Phys.\ Lett.} {\bf B 573} (2003) 217 [hep-th/0306166].

\bibitem{otha}N.~Ohta and J.~L.~Petersen,
{\it N=1 from N=2 superstrings,''
Phys.\ Lett.\ B {\bf 325}} (1994) 67
[arXiv:hep-th/9312187].
\bibitem{mukhi} L. Alvarez-Gaum\'{e}, D. Freedman and A. Mukhi, {\it The Background Field Method And The Ultraviolet Structure Of The Supersymmetric Nonlinear Sigma Model, Annals of Physics} {\bf 134} $\left( 1981\right) $ 85.
\bibitem{gates}
     S.J. Gates, Jr., {\it Supercovariant Derivatives, Super Weyl Groups, And N=2 Supergravity, Nucl. Phys.}  {\bf B 176} (1980) 397;
     S. J. Gates, Jr., W. Siegel, {\it Linearized N=2 Superfield Supergravity,  Nucl. Phys.} {\bf B 195} (1982) 39.
\bibitem{wit} B. de Wit and J. W. Van Holten, {\it Structure Of N=2 Supergravity, Nucl. Phys.} {\bf B 184} (1981) 77.
\end{thebibliography}
\end{document}